\documentclass[12pt]{article}
\usepackage{epsfig}
\usepackage{graphicx}  % standard LaTeX graphics tool
\usepackage{amssymb}

\begin{document}

\begin{center}
{\bf\Large Production of $B_c$ mesons via fragmentation in the
$k_T$-factorization approach}

\vspace{3mm}
{\sl V.A. Saleev}\footnote{Email: saleev@ssu.samara.ru} \\
{Samara State University, 443011 Samara, Russia}\\ and\\
{\sl D.V. Vasin}\footnote{Email: dmitriy.vasin@desy.de; vasin@ssu.samara.ru}\\
{II. Institut f\" ur Theoretische Physik, Universit\" at Hamburg, 22761 Hamburg,
Germany\footnote{On leave from Samara State University, 443011 Samara, Russia}}\\
%
%\authorrunning{V.A. Saleev and D.V. Vasin}
%

%\maketitle              % typesets the title of the contribution

\end{center}

\begin{abstract}

In the framework of the $k_T$-factorization approach we have
calculated in the fragmentation model the  $p_T$-spectra of $B_c$
mesons at the energies of the Tevatron and the LHC Colliders and
at the large $p_T$ domain. We compare the obtained results with
the existing experimental data and with the predictions obtained
in the collinear parton model.

\end{abstract}

\section{Introduction}

\hspace{0.7cm} Recently, $B_c$ mesons were observed
experimentally \cite{CDF1,CDF2}. These particles have a special
place in the doubly heavy meson family, because they consist of
quarks of different flavors.  The $B_c$ meson spectroscopy has
the same features as the spectroscopies of charmonia and
bottomonia, however the decay and production properties of $B_c$
mesons are unique. There is no annihilation
channel for $B_c$ mesons into hadrons via the gluons and, consequently, the lifetime
of the ground state of the $\bar b c$ system is large, given by
$\tau \approx 0.46$ ps \cite{CDF2}. The main channel of
$B_c$ meson decay occurs through the weak decays of the $b$-quark
or the $c$-quark, which are respectively about 25\% and 65\% of the total decay
width.

The $B_c$ meson hadroproduction cross section is approximately 100
times smaller than the ($b\bar b$)-pair production cross section, because
the two pairs of the heavy quarks are produced in a parton
subprocess. The mass spectrum and the decay properties of the $B_c$
mesons are studied in detail in the framework of potential quark
models \cite{QPM}, in the QCD sum rules methods \cite{SRQCD} and in the
OPE approach \cite{OPE}. The predictions at leading order in $\alpha_s$
for the $B_c$ meson production rates in the $\gamma \gamma$, $\gamma p$
and $pp$ interactions were obtained in the collinear parton model
\cite{Likhoded,Ruckl,Chang,Baranov,MS}.

The $B_c$ meson production cross section became large enough to
study experimental ($\sim 1$ nb) only in the region of very high
energy or in the region of very small $x\sim M/\sqrt{s}\sim
10^{-2}$,
where $x$ is the argument of the gluon distribution function
in a proton (the quark contribution to the hadronic production of
$B_c$ mesons is very small). In this region the $k_T$-factorization approach
\cite{Amp,KT} is more adequate for describing the perturbative evolution of the
gluon distribution function, which satisfies the BFKL \cite{BFKL} or CCFM
\cite{CCFM} evolution equations, in comparison with the collinear
parton model, which is based on the DGLAP equation \cite{DGLAP}.

In this Latter we have calculated $B_c$ meson
$p_T$-spectra at the energy range of the Tevatron and the LHC
Colliders in the fragmentation model and in the framework of the
$k_T$-factorization approach. In the fragmentation model we have
taken into consideration only the main contribution originating
from fragmentation of a b-quark into a $B_c$ meson \cite{Cheung}.

\section{The $k_T$-factorization approach}

In the $k_T$-factorization approach \cite{Amp,KT}, which generalizes
the collinear parton model to the region of small $x$,
the hadronic and partonic cross sections are related as follows:
\begin{eqnarray}
&&\sigma^{\mathrm{KT}}(p \bar p\to b \bar b X,
s)=\int{\frac{dx_1}{x_1}}\int{d|\vec
k_{1T}|^2}\int{\frac{d\varphi_1}{2\pi}} \Phi(x_1,|\vec
k_{1T}|^2,\mu^2)\times\nonumber\\&&\times\int{\frac{dx_2}{x_2}}
\int{d |\vec k_{2T}|^2}\int{\frac{d\varphi_2}{2\pi}}
\Phi(x_2,|\vec k_{2T}|^2,\mu^2)\hat\sigma (g^\star g^\star \to b
\bar b, \hat s)
%, |\vec k_{1T}|^2, \varphi_1, |\vec k_{2T}|^2, \varphi_2,
%\hat s)
\mbox{,}
\end{eqnarray}
where $\Phi(x,|\vec k_T|^2,\mu^2)$ is the unintegrated gluon distribution function
in a proton (unintegrated refers to the transverse momentum),
$|{\vec k}_{i,T}|^2$ is the virtuality of the initial reggeized gluon,
$k_{i}=x_{i} p_{i}+k_{i,T}$ is the 4-momentum of the initial
gluon, $k_{i,T}=(0,\vec k_{i,T},0)$ is the transverse momentum of
the initial gluon, $\varphi_i$ is the angle between the gluon
transverse momentum
and the fixed axis $OX$ in the plane $XOY$,
$\hat s=x_1 x_2 s-|\vec k_{1T}|^2-|\vec k_{2T}|^2$. In
the numerical calculations we have used the following
parameterizations for the unintegrated gluon distribution
function in a proton: JB by Blumlein \cite{JB}, JS by Jung
and Salam \cite{JS} and KMR by Kimber, Martin and Ryskin
\cite{KMR}.

To calculate the amplitude of partonic processes in the
$k_T$-factorization approach the polarization vector for the reggeized gluon
in the initial state is taken to be \cite{Amp,KT}
\begin{eqnarray}
\varepsilon^\mu(k_T)=\frac{k_T^\mu}{|\vec k_T|}\mbox{.}
\end{eqnarray}

The squared amplitude for one of the processes in $B_c$ meson production
in the $k_T$-factorization approach, namely
\begin{eqnarray}
g^\star+g^\star \to b + \bar b
\end{eqnarray}
was obtained earlier in \cite{Amp}. Here we use our original
results from Ref. \cite{SVpub}
\begin{eqnarray}
\overline{|M(g^\star+g^\star\to Q +\bar Q)|^2}=8 \pi^2 \alpha_s^2 (M_{11}+M_{22}+M_{33}+2 M_{12}+2
M_{13}+2 M_{23})\mbox{,}
\end{eqnarray}
where
\begin{eqnarray}
M_{11}&=&\frac{1}{3 {\tilde t}^2}
\Bigl(
\kappa_1 (4 \beta_1 - \kappa_1)(\kappa_1^2 + \kappa_2^2 - 4 \beta_2 \kappa_2)
+ 4 \lambda \beta_2 (\kappa_1 - 2 \beta_1)(4 \kappa_2^2 + {\tilde t} )
+ 4 \kappa_1^3 (\lambda \kappa_2 - \beta_1)
\nonumber\\&+&
4 \beta_2^2
(\kappa_1^2 + 4 \beta_1 \kappa_1 - 4 \lambda \kappa_1 \kappa_2)
+ 4 \lambda \kappa_2 (\lambda \kappa_2 - 2 \beta_1)(\kappa_1^2 -4 \beta_2^2)
- 4 \beta_1^2 (2 \beta_2 - \kappa_2)^2
+ \kappa_1^4
\nonumber\\&+&
4 \lambda \kappa_2 (2 \lambda \beta_2 + \beta_1)(2 \kappa_2^2 + {\tilde t})
+ 8 \lambda \beta_1 \kappa_2^3
+ 4 \lambda \kappa_2 ({\hat s} + {\tilde u})(\kappa_1 + \lambda \kappa_2)
+ {\tilde t} {\tilde u} \Bigr)\mbox{,}
\end{eqnarray}
\begin{eqnarray}
M_{22}&=&\frac{1}{3 {\tilde u}^2}
\Bigl(\kappa_2^4 - 4 \beta_2 \kappa_2^3  - 4 \beta_2^2 (\kappa_1^2+\kappa_2^2)
+ 16 \beta_1 \beta_2^2 (\kappa_1 - \beta_1)
+ 4 \lambda \bigl(4 \beta_1 \kappa_1^2  (\kappa_2 - 2 \beta_2 )
+{\tilde t} {\tilde u}
\nonumber\\&+&
\kappa_1 (\kappa_2^3 - 2 \beta_2 \kappa_2^2
+ (\kappa_2 - \beta_2)({\hat s} + {\tilde t} - 4 \beta_1^2)
- \beta_2 {\tilde u})
-\beta_1  (2 \beta_2 - \kappa_2) {\tilde u}\bigr)
- 16 \lambda^2 \beta_1^2 \kappa_1^2
\nonumber\\&+&
4 \lambda^2 \kappa_1 (4 \beta_1 \kappa_1^2  - \kappa_1 (\kappa_1^2 + {\tilde u})
+2 \beta_1  {\tilde u})
+\kappa_2 (4 \beta_2 - \kappa_2)((2 \beta_1^2 - \kappa_1)^2 + \kappa_2^2)
+ 4 \kappa_2^2 \beta_2^2
\Bigr)\mbox{,}
\end{eqnarray}
\begin{eqnarray}
M_{33}&=&\frac{3}{4 {\hat s}^2}
\Bigl((\kappa_1^2 +\kappa_2^2) {\hat s}
-4 (\beta_2 \kappa_1  - \beta_1 \kappa_2)^2
+ 4 \lambda \beta_1 \kappa_2(\kappa_1^2 - \kappa_2^2 - {\hat t} +
{\hat u})
\nonumber\\&-&
\lambda^2 \bigl(\kappa_1^4 + \kappa_2^4 - 8 \beta_1 \kappa_1^3
- 8 \beta_2 \kappa_2^3  - {\hat s}^2 + {\hat t}^2 +
8 \beta_1 \kappa_1  (\kappa_2^2
- 4 \beta_2 \kappa_2  + {\hat t} - {\hat u})
- 2 {\hat t} {\hat u}
+ {\hat u}^2
\nonumber\\&-&
2 \kappa_1^2 (\kappa_2^2 - 4 \beta_2 \kappa_2  - 8 \beta_1^2  + {\hat t} - {\hat u})
+ 2 \kappa_2^2 (8 \beta_2^2  + {\hat t} - {\hat u})
- 8 \beta_2 \kappa_2  ({\hat t} - {\hat u})\bigr)
\nonumber\\&-&
2 \lambda \bigl(8 \beta_2^2 \kappa_1 \kappa_2  + 2 \beta_2
(\kappa_1^3 - 4 \beta_1 \kappa_1^2  - 4 \beta_1 \kappa_2^2
- \kappa_1 (\kappa_2^2 + {\hat t} - {\hat u})) +
\kappa_1 \kappa_2 (8 \beta_1^2   - {\hat s})\bigr)\Bigr)\mbox{,}
\end{eqnarray}
\begin{eqnarray}
M_{12}&=&-\frac{1}{48 {\tilde t} {\tilde u}} \biggl(
4 \beta_1 \kappa_1^3
+ \kappa_2^2 ( {\hat s}- 8 \beta_1^2  - {\tilde t} - {\tilde u}
)
+ 32 \beta_1 \beta_2 (\beta_1 \kappa_2  - \beta_1 \beta_2)
+ {\hat s}^2
- {\tilde t}^2
- {\tilde u}^2
\nonumber\\&-&
\kappa_1^2 (\kappa_2^2 -\kappa_1^2
- 8 \beta_2 \kappa_2
+ 8 \beta_2^2  - 2 {\hat s})
+ 4 \beta_1 \kappa_1  (2 \kappa_2^2
- 8 \beta_2 \kappa_2
+ 8 \beta_2^2  - \kappa_1^2)
+
2 \lambda^2 \Bigl(\kappa_1^4
\nonumber\\&+& \kappa_2^4
- {\hat s}^2  +
{\tilde t}^2 + {\tilde u}^2 + 4 \beta_2 \kappa_2 {\tilde u}
+ 2 \kappa_2^2 {\tilde t} + 2 \kappa_1^2 ( 4 \beta_2 \kappa_2
+ {\tilde u})
+ 4 \beta_1 \kappa_1  (2 \kappa_2^2 - 4 \beta_2 \kappa_2
\nonumber\\&+&
{\tilde t})\Bigr) -
4 \lambda \Bigl(4 \beta_2^2 \kappa_2  (\kappa_1 - 2 \beta_1 ) -
\kappa_2 (-4 \beta_1^2 \kappa_1
+ \beta_1  (3 \kappa_1^2
+ \kappa_2^2 - {\hat s}) + \kappa_1 {\hat s})
- \beta_2  \bigl(\kappa_1^3
\nonumber\\&-&
4 \beta_1 \kappa_1^2
+ \kappa_1 (3 \kappa_2^2
+ 8 \beta_1^2  - {\hat s}) +
2 \beta_1  ({\hat s} -3 \kappa_2^2 - \kappa_1^2)\bigr)\Bigr)\biggr)\mbox{,}
\end{eqnarray}
\begin{eqnarray}
M_{13}&=&-\frac{3}{16  {\hat s} {\tilde t}}
\biggl(2 (\lambda^2-1) \kappa_1^4 + 2 \kappa_1^3
(2 \beta_1 - \lambda \kappa_2
- 3 \beta_1 \lambda^2
+ 3 \lambda \beta_2 ) +
2 \beta_1 \lambda  (2 \kappa_2^3
+ 4 \beta_2 \kappa_2^2
\nonumber\\&+&
4 \beta_2  ({\hat t} - {\hat u})
- \kappa_2 (16 \beta_2^2  - 2 {\tilde u})) -
\kappa_1^2 (2 (1 + \lambda^2) \kappa_2^2
- 2 ((5 \lambda^2 - 2) \beta_2
+ 6 \beta_1 \lambda) \kappa_2
\nonumber\\&+&
24 \lambda \beta_1 \beta_2  - 8 \beta_2^2  + 2 {\hat s} -
\lambda^2 {\hat s}
+ \lambda^2 {\tilde t} + 2 {\tilde u} - 3 \lambda^2 {\tilde u})
+ 2 \kappa_2 \beta_2  (
8 \beta_1^2
+ {\tilde t} + {\tilde u})
-  4 \beta_1^2 \kappa_2
\nonumber\\&-&
\kappa_2 {\tilde u}  +
\lambda^2 (4 \kappa_2^4 - 22 \beta_2 \kappa_2^3  - {\hat s}^2 + ({\hat t} - {\hat u})^2 -
\kappa_2^2 ({\hat s} - 32 \beta_2^2
- 5 {\tilde t} + 3 {\tilde u})
+ 2 \beta_2 \kappa_2  ({\hat s}
\nonumber\\&-&
7 {\tilde t} + 5 {\tilde u})) +
2 \kappa_1 (\beta_1 \lambda^2  (5 \kappa_2^2
- 16 \beta_2 \kappa_2  - 3 {\hat s} + {\tilde t} - 3 {\tilde u})
+ \beta_1  (4 \kappa_2^2 - 8 \beta_2^2
+ 2 {\hat s}
+ {\tilde t}
\nonumber\\&+&
{\tilde u}) +
\lambda (3 \kappa_2^3 - 11 \beta_2 \kappa_2^2  +
\kappa_2 ( 16 \beta_2^2 - 8 \beta_1^2
+ 2 {\tilde t}
- {\tilde u}) +
\beta_2  (16 \beta_1^2  + {\hat s} - 3 {\hat t} + 3 {\hat u})))\biggr)\mbox{,}
\end{eqnarray}
\begin{eqnarray}
M_{23}&=& - \frac{3}{16  {\hat s} {\tilde u}}
\biggl( \lambda^2 (4 \kappa_1^4
+ 2 \kappa_2^4 - 22 \beta_1 \kappa_1^3
- {\hat s}^2 +
2 \beta_1 \kappa_1  (5 \kappa_2^2
- 16 \beta_2 \kappa_2  + {\hat s} + 5 {\tilde t} - 7 {\tilde u})
\nonumber\\&-&
\kappa_1^2 (2 \kappa_2^2 - 16 \beta_2 \kappa_2
- 32 \beta_1^2  + {\hat s} + 3 {\tilde t} - 5 {\tilde u}) +
\kappa_2^2 ({\hat s} + 3 {\tilde t} - {\tilde u}) + ({\hat t} - {\hat u})^2
+ 8 \beta_2 \kappa_2 {\tilde u} )
\nonumber\\&-&
2 (\kappa_1^2 (\kappa_2^2 - 3 \beta_2 \kappa_2
+ 4 \beta_2^2  + {\tilde t}) +
\kappa_2 (\kappa_2^3 - \beta_2 \kappa_2^2  + \kappa_2 ({\hat s}
- 4 \beta_1^2  + {\tilde t})
+ \beta_2  (8 \beta_1^2  - {\hat s}))
\nonumber\\&-&
\beta_1 \kappa_1  (8 \beta_2^2 -2 \kappa_2^2
+ {\tilde t} + {\tilde u})) +
2 \lambda (\kappa_1^3 (\kappa_2 + 2 \beta_2 )
+ \beta_1 \kappa_1^2  (4 \beta_2 - 11 \kappa_2)
+ \beta_1  (4 \beta_2  ({\hat u} - {\hat t})
\nonumber\\&+&
3 \kappa_2^3 - 12 \beta_2 \kappa_2^2
+ \kappa_2 (16 \beta_2^2  + {\hat s} + 3 {\hat t} - 3 {\hat u}))
- \kappa_1 (3 \kappa_2^3 - 8 \beta_2 \kappa_2^2  -
\kappa_2 (16 \beta_1^2
- 8 \beta_2^2
- 2 {\hat s}
\nonumber\\&-&
3 {\tilde t} + {\tilde u}) +
\beta_2  (16 \beta_1^2  - {\tilde t})))\biggr)\mbox{,}
\end{eqnarray}
where
\begin{eqnarray}
{\tilde t} &=& {\hat t} - m_Q^2 \mbox{,}\\
{\tilde u} &=& {\hat u} - m_Q^2 \mbox{,}\\
\beta_1 &=& |{\vec p}_{T}| \cos{\varphi_1}\mbox{,}\\
\beta_2 &=& |{\vec p}_{T}| \cos{\varphi_2}\mbox{,}\\
\kappa_1 &=& |{\vec k}_{1T}|\mbox{,}\\
\kappa_2 &=& |{\vec k}_{2T}|\mbox{,}\\
\lambda &=& \cos{(\varphi_1-\varphi_2)}
\end{eqnarray}
and ${\vec p}_T$ --- transverse momentum of anti-quark. Since
these are more suitable for numerical calculations. In those
calculations, which were in the framework of the collinear parton
model, we used the GRV \cite{GRV} parameterization for the gluon
distribution function.

\section{The fragmentation model}

The analysis of the $B_c$ meson gluon-gluon production
in the framework of the collinear parton model shows
 \cite{Likhoded,Ruckl,Chang} the dominant role of the fusion mechanism up to
$p_T\thickapprox 30$ GeV. It takes into account the total gauge
invariant set of the diagrams which describes the parton process
\begin{eqnarray}
g+g \to B_c + b + \bar c\mbox{.} \label{eq:ggBcbc}
\end{eqnarray}
Thus, the fragmentation approximation, incorporating factorization
of the heavy quark production process and the soft process of a meson created in the final state,
applies only at $p_T > 30$ GeV. Of course, at the Tevatron Collider this
region of $p_T$ is far from being available for experimental study. On the other hand, at the
LHC Collider the $p_T\le 50$ GeV region will be under consideration.

In the fragmentation model, the $B_c$ meson production cross
section can be presented as follows:
\begin{eqnarray}
d \sigma(pp\to B_c X)=\sum_i \int{dz} D_{i \to B_c}(z)\cdot d \hat
\sigma(pp\to i)\mbox{,}
\end{eqnarray}
where the sum is taken over all the relevant partons $i=c$, $b$
and  $g$ (but the main contribution comes from $b$-quark
fragmentation).

The fragmentation functions for the $b$-quark splitting into a
vector $B_c^*$ and pseudoscalar $B_c$ meson at the starting scale
of the perturbative QCD evolution were obtained in Ref.
\cite{Braaten,Likhoded-z0}:
\begin{eqnarray}
D_{\bar b\to B_c}(z,\mu_0)&=&\frac{8 \alpha_s^2(\mu_0)
|\Psi_{B_c}(0)|^2}{81 m_c^3} \frac{r (1 - z)^2 z}{(1 - (1 - r)
z)^6} \Bigl(6 - 18 (1 - 2 r) z
+ (21 - 74 r + 68 r^2) z^2 \nonumber\\
&-&2 (1 - r) (6 - 19 r + 18 r^2) z^3
+3 (1 - r)^2 (1 - 2 r + 2 r^2) z^4\Bigr)\mbox{,}\\
D_{\bar b\to B_c^\star}(z,\mu_0)&=&\frac{8 \alpha_s^2(\mu_0)
|\Psi_{B_c}(0)|^2}{27 m_c^3} \frac{r (1 - z)^2 z}{(1 - (1 - r)
z)^6} \Bigl(2 - 2 (3 - 2 r) z + 3 (3 - 2 r + 4 r^2) z^2
\nonumber\\&-& 2 (1 - r) (4 - r + 2 r^2) z^3 + (1 - r)^2 (3 - 2
r + 2 r^2) z^4\Bigr)\mbox{,}
\end{eqnarray}
where $r=\frac{m_c}{m_c+m_b}$. $|\Psi_{B_c}(0)|^2$ is the squared
$B_c$ meson wave function at the origin, which can be calculated
using the nonrelativistic potential quark model \cite{Psi0}. In the numerical calculations, we
used the result
$\Psi_{B_c}(0)=\sqrt{\frac{m_{B_c}}{12}}f_{B_c}$, where
$f_{B_c}=490-560$ MeV is the $B_c$ meson leptonic decay constant.

The QCD evolution of the fragmentation functions $D_{\bar b\to
B_c}$ and $D_{\bar b\to B_c^\star}$ are described by the DGLAP
\cite{DGLAP} evolution equation
\begin{eqnarray}
\mu^2\frac{\partial D}{\partial
\mu^2}(z,\mu^2)=\frac{\alpha_s(\mu^2)}{2\pi}\int_z^1
\frac{dx}{x}P_{q\to q}(\frac{x}{z}) D(x,\mu^2)\mbox{,}
\end{eqnarray}
where $P_{q\to q}(z)$ is the standard quark-quark splitting
function.

\section{The results}

\begin{figure}[t!]
\begin{center}
\includegraphics[width=1.0\textwidth, clip=]{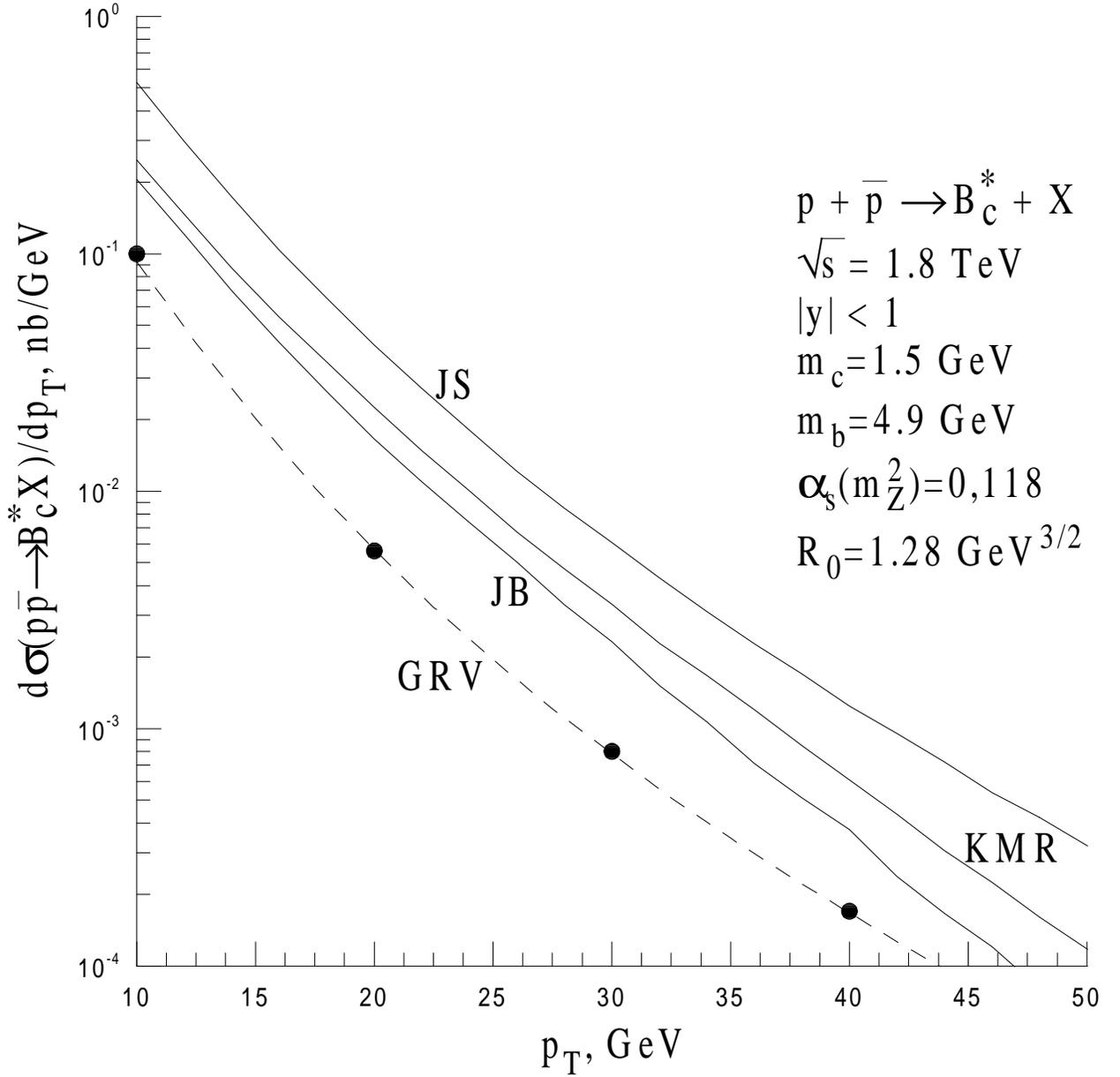}
\end{center}
\caption[]{The  $B_c^\star$
meson $p_T$-spectrum at $\sqrt{s}=1.8$ TeV and $|y|<1$ in
the fragmentation model. The continuous lines are a result of
calculations in the $k_T$-factorization approach with the
different parameterizations of the unintegrated gluon distribution
function. The dashed line is the prediction from the collinear
parton model. The points show the results obtained in the Ref.
\cite{Cheung}. \label{fig:Tev_FR}}
\end{figure}

\begin{figure}[t!]
\begin{center}
\includegraphics[width=1.0\textwidth, clip=]{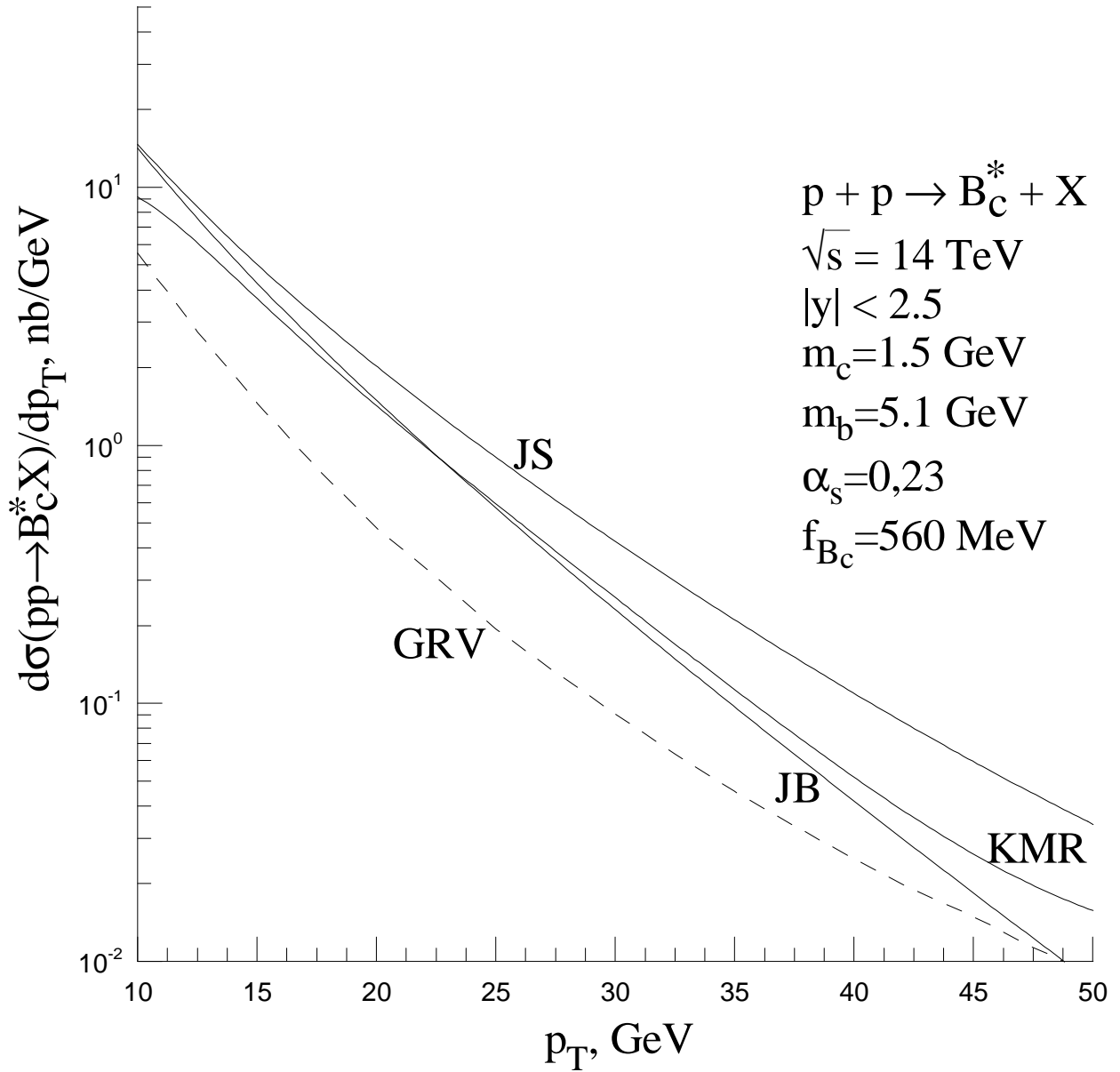}
\end{center}
\caption[]{The $B_c^\star$ meson $p_T$-spectrum at $\sqrt{s}=14$ TeV and $|y|<2.5$ in the
fragmentation model. The continuous lines are results of
calculations in the $k_T$-factorization approach with the
different parameterizations of the unintegrated gluon distribution
function. The dashed line is the prediction from the collinear
parton model. \label{fig:LHC_FR}}
\end{figure}

The results of our calculation for the $p_T$-spectra of the
$B_c^\star$ meson are shown in Fig. \ref{fig:Tev_FR} for
the energy of the Tevatron Collider $\sqrt{s}=1.8$ TeV, and in the
Fig. \ref{fig:LHC_FR} for the energy of the LHC Collider
$\sqrt{s}=14$ TeV. The spectra that we obtain are compared with those
obtained earlier in the collinear parton model via the
fragmentation mechanism. The curves in Fig. \ref{fig:Tev_FR}
correspond to a choice of the parameters from Ref. \cite{Cheung}:
$m_c=1.5$ GeV, $m_b=4.9$ GeV, $\alpha_s = \alpha_s(p_T^2+ m_{B_c}^2)$ and
$f_{B_c}=490$ MeV. The curves in Fig. \ref{fig:LHC_FR} were
obtained using the set of parameters from
Ref.\ \cite{Likhoded}: $m_c=1.5$ GeV, $m_b=5.1$ MeV, $\alpha_s
\simeq 0.23$ and $f_{B_c}=560$ MeV.

Fig. \ref{fig:Tev_FR} shows that our result obtained in the
collinear parton model agrees well with the result obtained earlier in
Ref. \cite{Cheung}. The curves obtained in the
$k_T$-factorization approach lie approximately 2---5 times higher than
the collinear parton model prediction in the region $p_T> 10$ GeV.
The maximum value of the $B_c$ meson production cross section is obtained
using the JS \cite{JS} parameterization of the unintegrated gluon distribution
function, and the minimum value is obtained using the JB \cite{JB} parameterization.

Note that the slopes of the $p_T$-spectra obtained in the
collinear parton model and in the $k_T$-factorization approach are
equal, and the situation is the same as in the case of
$D^\star$ meson production in $ep$ interactions
\cite{SV,ZBJ}.

At present, there is no experimental information on the $p_T$-spectra of $B_c$
mesons. It is known that the integrated production
cross section in the region $p_T > 6$ GeV, $|y| < 1$ and $\sqrt{s}=1.8$
TeV is given by $\sigma_{B_c} \simeq 10 \pm 6$ nb \cite{CDF1,CDF2}. As
shown in Ref. \cite{Likhoded}, the uncertainties in the
calculations are about $\sim 50$\%. This value depends on the
choice of the masses, the $\alpha_s$ constant and the leptonic
decay constant $f_{B_c}$. Furthermore, our calculations performed in the
$k_T$-factorization approach show that
the $B_c$ meson production cross section has a strong dependence  on the choice of the
unintegrated gluon distribution function --- a variation by a factor of 2 was observed.

The obtained values of the integrated $B_c$ and $B_c^\star$ meson
production cross section at the energies of the Tevatron and the
LHC Colliders are presented in Table \ref{tab:1}. One can
see that, with the fragmentation model, the results obtained in the
collinear parton model  are smaller to those obtained in the $k_T$-factorization approach.
The theoretical predictions are shown together with their
uncertainties, which are significantly larger in the
$k_T$-factorization approach. The absolute value of the cross
section obtained in the collinear parton model
($\sigma_{B_c}=1.7\pm 0.8$ nb) is much smaller than the CDF
experimental data ($\sigma_{B_c}=10\pm 6$ nb) \cite{CDF1,CDF2}.
On the other hand, the prediction of the $k_T$-factorization approach
agrees well with the data ($\sigma_{B_c}=7.4\pm 5.4$ nb).

Taking into account the relative roles of the fusion and
fragmentation mechanisms in $B_c$ meson hadroproduction, one can
suppose that the cross section calculated in the fusion model and
in the $k_T$-factorization approach will be more large than that obtained
in the fragmentation model and in the $k_T$-factorization
approach. The calculation of the $B_c$ meson hadroproduction
rates via the partonic subprocess (\ref{eq:ggBcbc}) with the reggeized initial
gluons will be presented in our forthcoming paper.

\section{Acknowledgments}

We thank  B.~Kniehl and  A.~Likhoded for useful discussions about
the results obtained. The work is supported by the Russian Federal
Agency of Education under Grant A04-2.9-52. The part of the D.V. work was 
done in the framework of the Grant "Mikhail Lomonosov", which is supported by
DAAD and by Russian Ministry of Education.
D.V. thanks the International Center of Fundamental Physics in Moscow and
Dynastiya Foundation for the financial support received while this work was
done.

\begin{table}
\caption{The summed $B_c$ and $B_c^\star$ meson
production cross sections in the different models at the energies
of the Tevatron and LHC Colliders. The data from the CDF
Collaboration\cite{CDF1,CDF2} are shown. The cross sections are in nb.}
\bigskip
\begin{tabular}{|c|c|c|c|c|c|}
\hline
$\sqrt{s},$ & $|y|$ & $p_{T,min}$, & Parton & $k_T$-factorization& Experimental \\
 TeV &  & GeV &  model & approach & data
 \\\hline $1.8$ & $< 1$ & $6$ & $1.7 \pm 0.8$ & $7.4
\pm 5.4$ & $10 \pm 6$
\\\hline
$14$ & $< 2.5$ & $10$ & $28 \pm 14$ & $122 \pm 90$ & ---
\\[1mm]
\hline
\end{tabular}
\label{tab:1}
\end{table}


\end{document}